\documentclass[12pt]{book}

\usepackage[dvips]{graphicx,color}
\usepackage{makeidx,observe,cosmology}

\makeauthorindex
\makeindex

\BookTitle{New Trends in Theoretical and Observational Cosmology}
\CopyRight{\copyright 2001 by Universal Academy Press, Inc.}

\begin{document}

\BookTitle{\itshape New Trends in Theoretical and Observational Cosmology}
\CopyRight{\copyright 2001 by Universal Academy Press, Inc.}

\pagenumbering{arabic}

\chapter{Measurements of The CMB: Current Status and Future Prospects}

\author{Lyman Page\\
{\it Dept. of Physics, Princeton University, Princeton, NJ 08540-0708}}

\AuthorContents{L.\ Page} 

\AuthorIndex{Page}{L.}

\section*{Abstract}

The study of the Cosmic Microwave Background (CMB) is a rapidly maturing
field. In remarkable advances in the past two years, experiments 
have begun to probe the
detailed structure of the CMB angular spectrum thereby providing insight
into the contents of the universe. We briefly review the BOOMERanG and
DASI results with an emphasis on what the data tell us with only
minimal cosmological assumptions. After a short discussion of a
grand future for CMB physics, key features of the {\sl MAP} satellite, 
now in orbit, are described. We also discuss how we anticipate 
{\sl MAP} will advance the field beyond what we know now.

\section{Introduction}
This is a very exciting time for those studying the CMB. The decades of work
that have gone into developing sophisticated technologies and
observational techniques are beginning to pay off. A number of teams now
have measured features in the CMB angular spectrum, measured its
frequency spectrum, and have deduced cosmological parameters that are in
rough accord with what one obtains with other methods based on entirely
different physics.

The CMB is a particularly potent probe of cosmology because its
characteristics may be computed to high accuracy and it may be 
measured to high accuracy. As long as the systematic errors associated
with the experimental technique are small, the detected signal has
direct cosmological import. Potentially confounding emissions 
from our galaxy and 
sources between us and the surface of last scattering are small.
To first order, such ``foregrounds'' may be ignored at many angular
scales and frequencies\cite{te96}.
To the extent the foregrounds do affect a result, they may be measured and
subtracted: the CMB frequency spectrum is unique. 

\section{Current Status}

The past two years have seen wonderful measurements
of the CMB anisotropy. The first peak, the position and amplitude
of which were known within the CMB community in the last millennium
(e.g., \cite{bartlett98,bahcall99,dk00,kp00,hu00}), 
has now been made manifestly apparent in high signal to noise maps 
by the BOOMERanG \cite{net01} and and MAXIMA teams \cite{Hanany00}.
The location of this peak
is a measure of the angular diameter of the sound horizon when the
photons first decoupled from the primordial baryons. The sound
horizon gives a physical length scale in the early universe which, 
from our vantage,
is affected primarily by the geometry of space. In a flat geometry, its 
angular diameter is $\approx 1/2^{\circ}$; in a closed geometry,
it will appear larger and in an open geometry smaller 
(e.g.,\cite{doris78,kam94,Bond94}). The location
of the peak at $l\approx 215$, as shown in Figure 1, 
indicates that the geometry of the universe is parsimoniously 
described as flat.

\begin{figure}[t]
  \begin{center}
    \includegraphics[height=9pc]{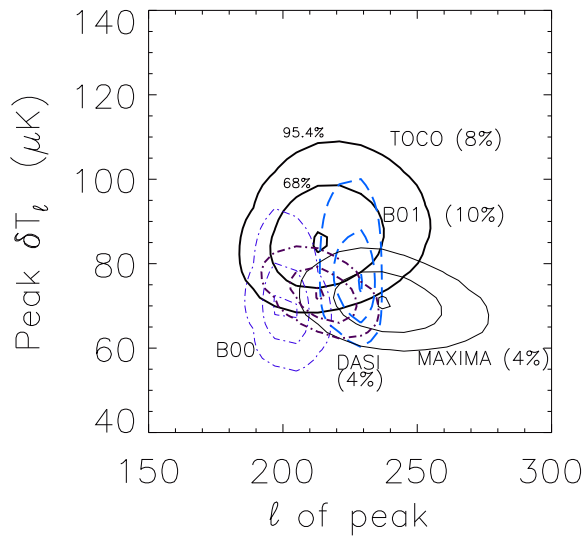}
  \end{center}
\vspace{1.0truein}
\caption{The position and amplitude of the first peak 
following the Gaussian temperature method \cite{kp00}. 
For TOCO\cite{Miller99}  
$l_{peak}=216\pm14$ and $\delta T_l=86\pm8~\mu$K.
The preliminary BOOMERanG\cite{dB00} data give 
$l_{peak}=201$ and $\delta T_l=70~\mu$K; 
the reanalyzed and expanded data set \cite{net01} gives $l_{peak}=226$
and $\delta T_l=77~\mu$K. (The North American Flight is omitted.) 
The BOOMERanG beam uncertainty is not accounted for; it will
tend to broaden the distribution in $l$. MAXIMA\cite{Lee01} yields 
$l_{peak}=238$ and $\delta T_l=71~\mu$K and DASI \cite{Halverson01} 
gives $l_{peak}=213$ and $\delta T_l=74~\mu$K.
Calibration uncertainty has been taken into account though the 
correlations between
bands have not. When treated consistently, TOCO, B01, MAXIMA, and
DASI pick out values for the peak position and amplitude that
are within $2\sigma$ of each other. For all experiments the angular
spectra for the peak fits are normalized at $l=10$ to DMR\cite{kp00}; 
this is most important for DASI and less
important for the other experiments. From  Miller {\it et al.} 2001.
\cite{Miller01}}
\end{figure} 

This is a triumph for physical cosmology. The mechanisms leading to a
peak have been known for over thirty years and detailed models
of the shape predated all detections of the anisotropy. As the 
net matter density inferred from the distribution and velocities
of galaxies is about 1/3 the critical density, there must be 2/3
the critical density in some new form of energy. This conclusion is 
independent, though similar, to that drawn from the supernovae results 
as discussed at this conference by Saul Perlmutter.     

\begin{figure}[t]
  \begin{center}
    \includegraphics[height=20pc]{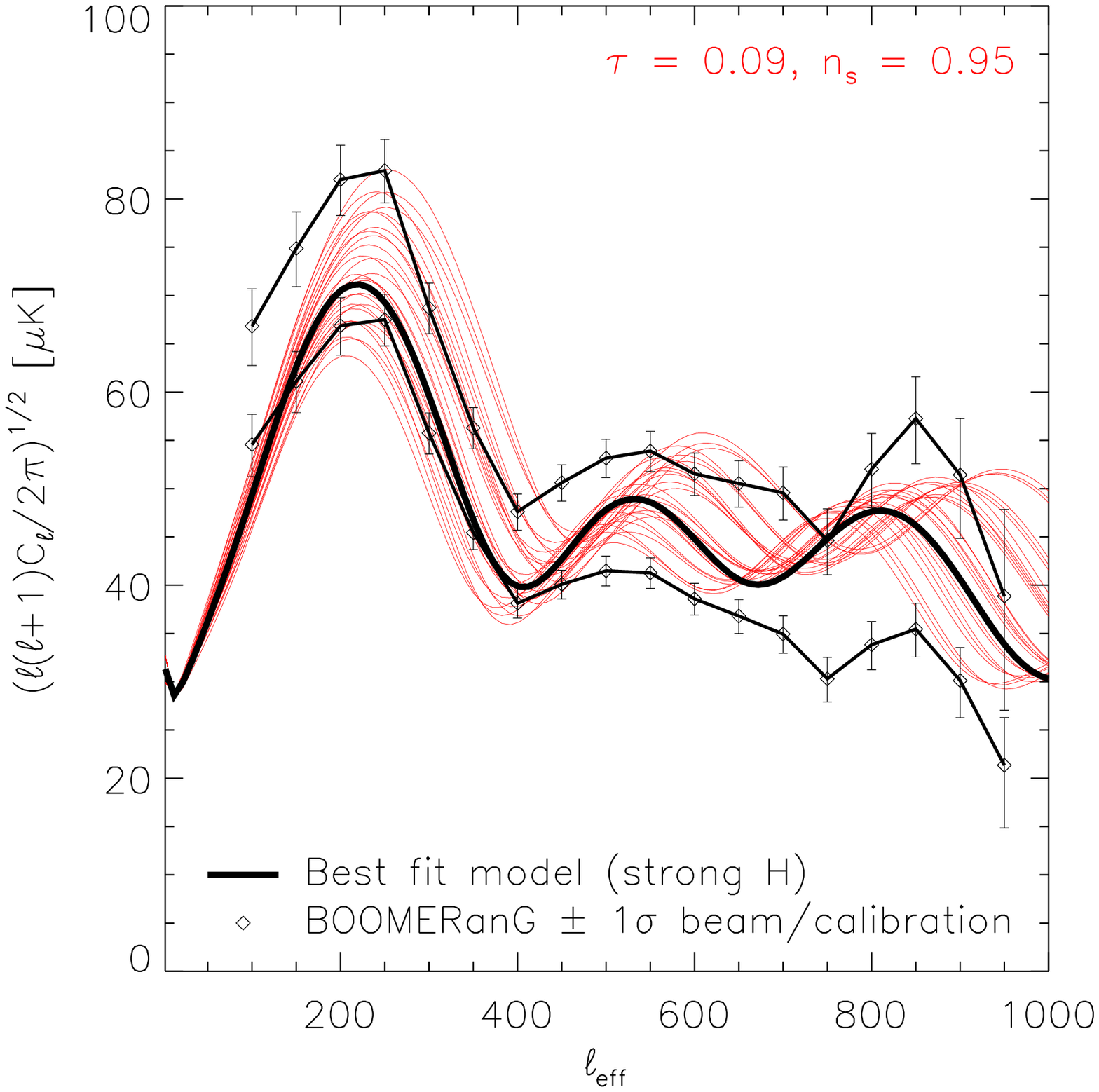}
  \end{center}
  \caption{The BOOMERanG data\cite{net01}. The top and bottom
curves show the $1\sigma$ errors for the beam and calibration
uncertainties added in quadrature. 
Assuming a flat prior ($\Omega=1$), the best adiabatic model fit
to the data is
$(\Omega_b h^2,\Omega_c h^2, \Omega_\Lambda,n_s,\tau)
= (0.021\pm0.003,0.13\pm0.04,0.57^{+0.12}_{-0.37},
0.95^{+0.09}_{-0.08},0.09^{+0.13}_{-0.07})$, shown in black.
The set of models shown in red are derived from the best fit,
allowing the parameters to vary by $\pm1\sigma$. The tilt $n_s$ and
depth to reionization $\tau$ are kept fixed because their combined
action is similar to an overall scale factor. From M. Nolta.} 
\end{figure}

\begin{figure}[ht]
  \begin{center}
    \includegraphics[height=20pc]{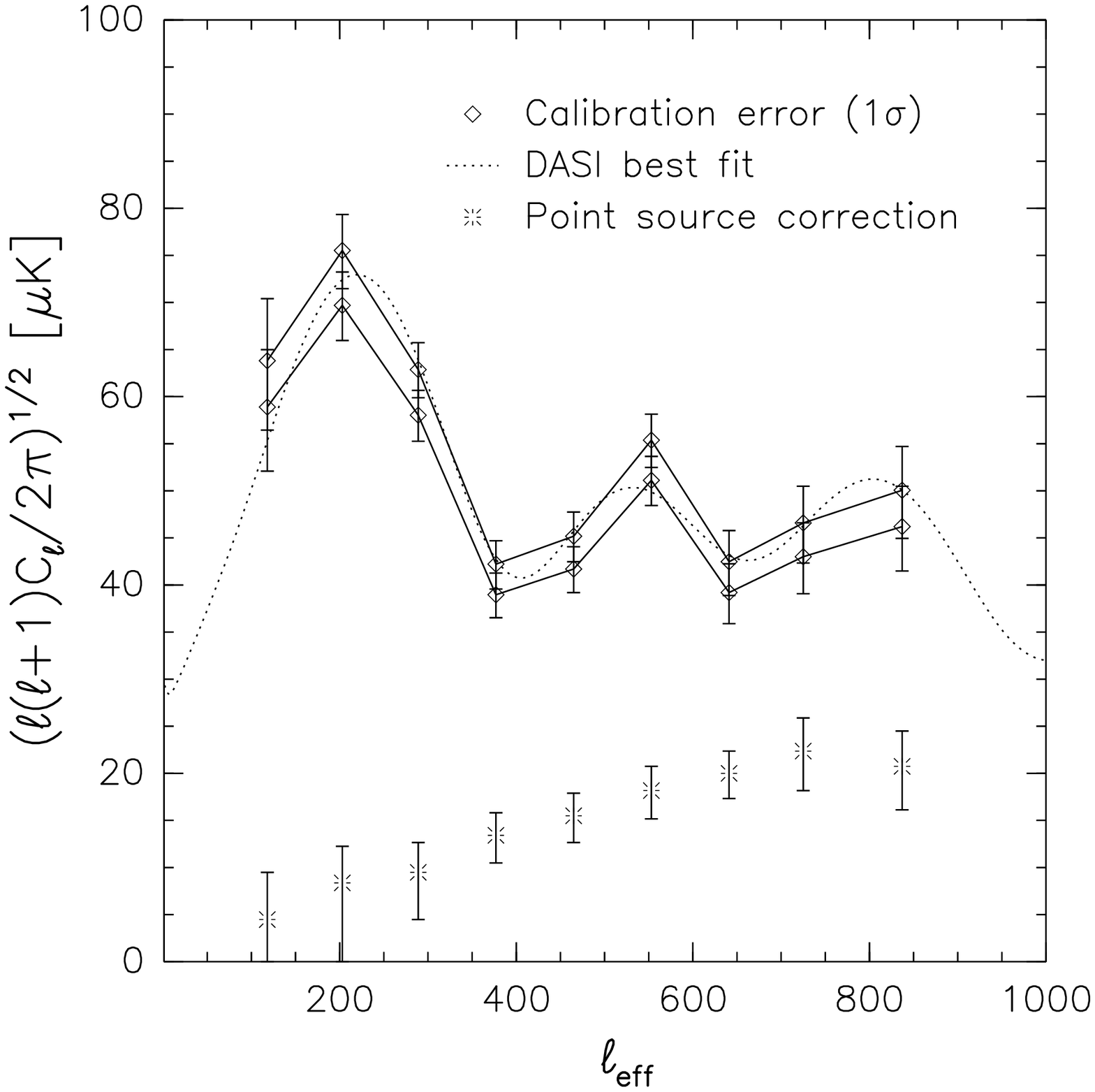}
  \end{center}
  \caption{The DASI data\cite{Pryke01}. The upper and lower
lines through the data points show the calibration error.
The points on the bottom show the point source contribution
to the CMB data. They add in quadrature and so are a smaller fraction
of the CMB signal than a direct read of the plot would indicate. The dotted
line shows their best model. From M. Nolta}
\end{figure}  

Many of the models that predicted the first peak also predict 
that there should be
higher order peaks or ``acoustic oscillations.'' The
second peak, at $l\approx 400$, is particularly sensitive to 
the baryon density.
More baryons lead to more inertia in the primordial plasma, in turn
leading to relatively more pronounced compression peaks ($1^{\rm st}$
and $3^{\rm ed}$) than rarefaction peaks ($2^{\rm nd}$ and $4^{\rm th}$).
Though there were early indications from BOOMERanG that a second peak 
might be suppressed\cite{dB00}, a more complete analysis 
that included much more data\cite{net01} 
showed evidence of it. Independently, the DASI experiment
saw the drop off from the first peak and also showed evidence for
the second peak. The data from these experiments, along with calibration
and beam errors, is shown in Figures 2 \& 3.  

The significance of the detection of the second peak depends on one's 
prior assumptions and analysis method\cite{Miller01b,Durrer01}.
A combination of all the data by Wang, Tegmark, 
and Zaldarriaga\cite{Wang01} shows clear evidence for it. A non-believer
might point to the fact that $\chi^2/\nu$ for a fit to a flat line
for the  $l>400$ data, is 23/11 for BOOMERanG (correlations not published)
and is 8.66/4 for DASI (correlations included).
In other words, random noise would produce the same fluctuations
away from the $l>400$ mean in 5\% of the cases 
for BOOMERanG (from an analysis with correlations \cite{dB01}) and in 
7\% of the
cases for DASI. The third peak is on less firm footing than the second.
 
Extracting cosmological parameters from the CMB anisotropy 
started soon after its discovery\cite{Smoot}. The current CMB 
data give solutions for 
roughly three to four cosmological eigenparameters 
(e.g., \cite{Bond94,efs01}) with
similar results arising from just the BOOMERanG data or from all 
data combined. An analysis of all data \cite{Wang01} gives 
$\Omega_b h^2=0.02^{+0.06}_{-0.01}$, $\Omega_m h^2=0.1^{+0.07}_{-0.06}$,
$\Omega_\Lambda = 0.6^{+0.32}_{-0.55}$, $n_s=0.93^{+0.75}_{-0.16}$
( 2$\sigma$ ) though the errors are correlated. 
It is worth keeping in mind that the ``lambda adiabatic CDM''
model is assumed for all of the parameter extraction. 

\section{Future Prospects}

The future of CMB observations is bright. The recent successes have
given us just a glimpse of what may be learned. The {\sl MAP} satellite,
which is discussed in greater detail below, is in orbit and taking
data. The Planck 
satellite, with higher resolution and greater sensitivity 
than {\sl MAP}, is planned for 
a 2007 launch. The frontiers after the primary anisotropy is mapped out
include measuring the polarization and, through the ``secondary
anisotropies,'' the onset of cosmic structure formation. A host of
balloon and ground based experiments are planned for making 
the many observations for which space is not necessary
and for laying the groundwork for future space missions.

If the currently most favored class of models holds, the polarization in
the CMB, through it dependence on gravity waves, will give us
information about fundamental fields in the $t<10^{-20}~$s universe.
It may possibly distinguish between such alternatives as inflation
and a cyclic universe\cite{PJSNT01}. As the observable signal 
is very small, and may be masked by foreground effects, a great deal 
of work is left to do. Already though, researchers are discussing a 
dedicated satellite (CMBPOL) for measuring the polarization. 
 
\section{\sl MAP}

\begin{figure}[t]
  \begin{center}
    \includegraphics[height=22pc]{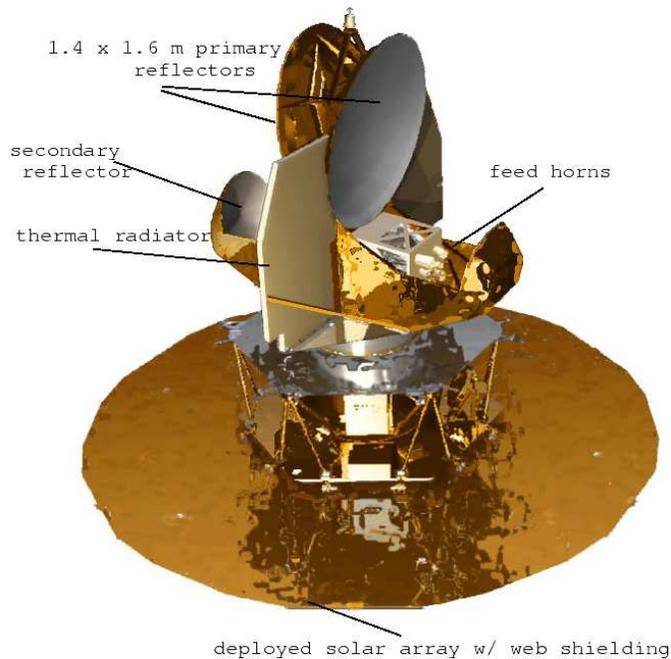}
  \end{center}
\caption{Picture of {\sl MAP} with the solar arrays deployed. The receivers
are located directly beneath the primary reflectors and are passively
cooled by the thermal radiators. The Earth, Moon, and Sun are beneath 
the solar arrays.}
\end{figure}

The next large influx of CMB data is expected to come from the {\sl MAP}
satellite. The primary goal of {\sl MAP}
is to produce a high fidelity, polarization sensitive map of the full
sky. {\sl MAP} was successfully launched
on June 30, 2001 at 3:46:46 from Kennedy Space 
Flight Center. The launch by
Boeing was within 183 ms of the opening window! 
So far, the
mission has proceeded as planned and the instrument performance is nominal. 

Data from {\sl MAP} will stand apart from previous data sets in a number
of ways:
\begin{itemize}
\item {\sl MAP} will produce all sky maps in five frequency bands
with more than 45 times the sensitivity of {\sl COBE/DMR} and
30 times the angular resolution.
\item {\sl MAP}'s maps, to a good approximation, will have the same 
relative temperature 
errors on large angular scales ($>90^{\circ}$) as on small 
ones ($<1^{\circ}$). To extract 
cosmological parameters, current experiments pin the $C_l$ spectrum at
low $l$ to {\sl COBE}.
\item Systematic uncertainties will be controlled at new levels.
Calibration and/or beam uncertainty dominate all current measurements and
there may be lower level systematic errors of which we are not aware.
{\sl MAP} will be calibrated on the CMB dipole to better than 1\% and the 
beams will be mapped in flight multiple times. Because of the long time
spent in a very stable orbit, numerous other consistency and systematic
checks are possible.
\item True maps will be produced with small pixel-to-pixel 
correlations. The fidelity of a map to the underlying true flux 
distribution depends on the level of filtering in going from the time
stream to the spatial map. To a very good approximation, 
{\sl MAP} senses radiation from only the direction in which it is pointing
and the reconstruction from the differential data to a map constrains
artifacts at the sub-$\mu$K level. Only the
slightest filter is applied. Up to an unknown single additive 
constant, what you see in the map will be what's in the sky. 
\end{itemize} 

\begin{figure}[t]
  \begin{center}
    \includegraphics[height=18pc]{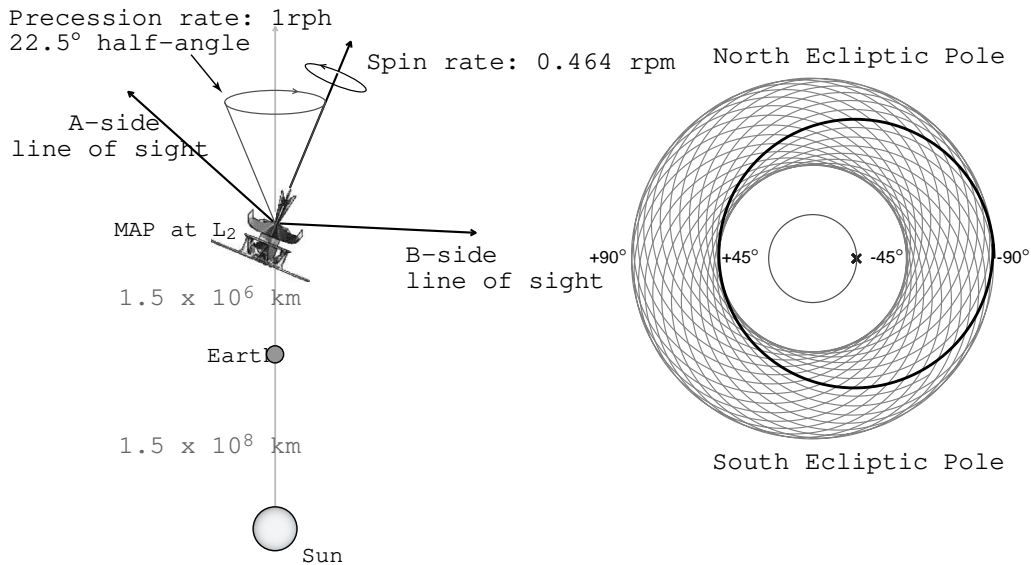}
  \end{center}
\caption{{\sl MAP}'s scan pattern from L2.
The dark circle on the left hand drawing depicts the path covered by two
beams for one rotation; the innermost circle is the path of the spin
axis during one precession. The four modulation time scales are: 2.5 kHz
switching between the two sides, 2.1 min spin period, 1 hr precession to
cover $\approx 30$\% of sky, and 6 months to cover the whole sky.}
\end{figure}
 
\subsection{Essential components of the design}

{\sl MAP} is a differential microwave radiometer.
Table 1 gives the approximate characteristics of the instrument
though the flight performance may differ slightly.
The components of the {\sl MAP} mission---receivers, optics, scan strategy,
thermal design, and attitude control---all work together. Without any
one of them, the mission would not achieve its goals.
One guiding philosophy, however, is that a differential measurement
with a symmetric instrument is highly desirable as discussed, for
example, by Dicke (1968). The reason is that in such a system 
the receivers outputs are, to
first order, insensitive to changes in the satellite temperature
or other radiative properties. 

\begin{table}[htb]
\begin{center}
\begin{tabular}{lccccc}
\hline 
Band & $f_{\rm center}$ (GHz)& Bandwidth (GHz) & \# Signals 
& $\theta_E$ (deg)& $\theta_H$ (deg) \\
\hline
K & 23 & 5.3 & 2 & 0.95 & 0.75 \\
${\rm K_a}$ & 33 & 7 & 2  & 0.7 & 0.6  \\
Q & 41 & 8.4 & 4 & 0.45 & 0.5  \\
V & 61& 12 & 4  & 0.3  & 0.35  \\
W & 95& 17 & 8  & $<0.25$  & $<0.25$   \\
\hline
\end{tabular}
\caption{ {\sl MAP} Frequencies, Noise Bandwidths, and Beam Sizes.
The beams are elliptical and so both E and H plane Gaussian
FWHM are given. There is one pair of feeds in both K and K${\rm _a}$
bands. The two signals
correspond to two polarizations. The sensitivity for each band will be
$\approx 35~\mu$K for each $3.2\times10^{-5}$~sr pixel 
for two years.}
\end{center}
\end{table}

{\sl MAP} uses a pair of composite back-to-back offset shaped Gregorian
telescopes that focus celestial radiation onto ten pairs of back-to-back
corrugated feeds. The primary optical axes of the two telescopes are
separated by $141^{\circ}$. Separate feeds accept radiation in five
frequency bands between 20 and 100 GHz.  Two orthogonal polarizations
are selected by an orthomode transducer at the base of each feed.

There are two keys to making large area maps with negligible pixel-to-pixel
correlations: (a) stable receivers and (b) a highly interlocking scan 
strategy. One measure of the stability of the receivers 
is its ``$1/f$'' knee. This is the frequency at which the power spectral
density (PSD) of the receiver output is double that of the value at a high
frequency. Laboratory tests showed that the $1/f$ knee is 
$<10~$mHz and that the PSD from 20~mHz to 3~kHz is white. 
The stability of the test setup limited the low frequency performance;
the receivers will be more stable in space. 
 
The receiver stability is due to a number of factors. Foremost is the
design. {\sl MAP} uses NRAO HEMT amplifiers in a correlation
configuration with 2.5~kHz phase switching. Thus, the detector output
is proportional to the difference in polarized flux at 2.5~kHz from
two celestial pixels separated by $\approx 141^{\circ}$. Second is
the thermal stability. The passive 
thermal design combined with a constant Sun angle and an L2 orbit 
$1.5\times10^{6}$~km from Earth results in a thermal drift rate
measured in mK/hr, unattainable from any platform near Earth.

With the highly interlocking scan shown in Figure 5, the value in
each pixel on the sky is referenced to that in hundreds of other
pixels. As the cross-linking occurs on multiple time scales, 
there is a highly selective spatio-temporal filter that links 
the time ordered differential
data to an all sky map. There are a number of proven 
algorithms for processing the data \cite{Wright95,Hivon01} 
that are used by the {\sl MAP} team.

\subsection{Science from {\sl MAP}}
There is still a large gap between the models that are
currently under consideration and the possible models
that can describe the CMB. Despite the recent
observational advances, the favored model of the late 1990's is still 
the most favored model\cite{Lange01} though our confidence in it has 
been considerably enhanced. Only with a map with negligible systematic 
artifacts can different models be tested. 
The following list is a sample of the anticipated science:
 
\begin{itemize}
\item CMB. {\sl MAP} will measure the CMB angular spectrum 
to $l\approx1000$ with a cosmic variance limit at $l\approx 500$,
greatly extending {\sl COBE}'s of  $l\approx 20$.
A simulation of the angular spectrum is shown in Figure 6. 
With the data one can test for Gaussanity, polarization, and 
the polarization-temperature cross-correlation. {\sl MAP} can tell 
if there are significant isocurvature or tensor modes or if just 
six parameters describe the $\approx 10^6$ data points.

\item Large scale correlations of other surveys with the CMB. Is there a
SZ decrement in the supergalactic plane?\cite{sb98} Are there
correlations with point sources due to ``dark energy'' 
or quintessence?\cite{sb01}. 

\item ISM. Is there
significant emission from spinning dust?\cite{dl99}, Does a $10^{4}$K plasma
describe all the free-free emission? Is dust naturally associated with 
free-free/H$_{\alpha}$ emission? What is the polarization of the ISM?

\item Non-linear structures. {\sl MAP} should detect the SZ effect in the
largest ten or so clusters \cite{ar00} and there should be 
a detectable lensing signal from a cross correlation with the Sloan
Digital Sky Survey \cite{peiris00}.
 
\item Sources. {\sl MAP} will give a full sky census of bright 
extragalactic sources. It will shed light on their
emission properties between 30 and 100 GHz and will be sensitive to 
new classes of sources.

\item Calibration. One of the largest systematic errors in many
experiments is calibration. {\sl MAP} will calibrate celestial sources to 
the Earth's velocity through the CMB dipole. This enables
the precise comparison between multiple cosmic probes.
 
\end{itemize}

\begin{figure}[t]
  \begin{center}
    \includegraphics[height=20pc]{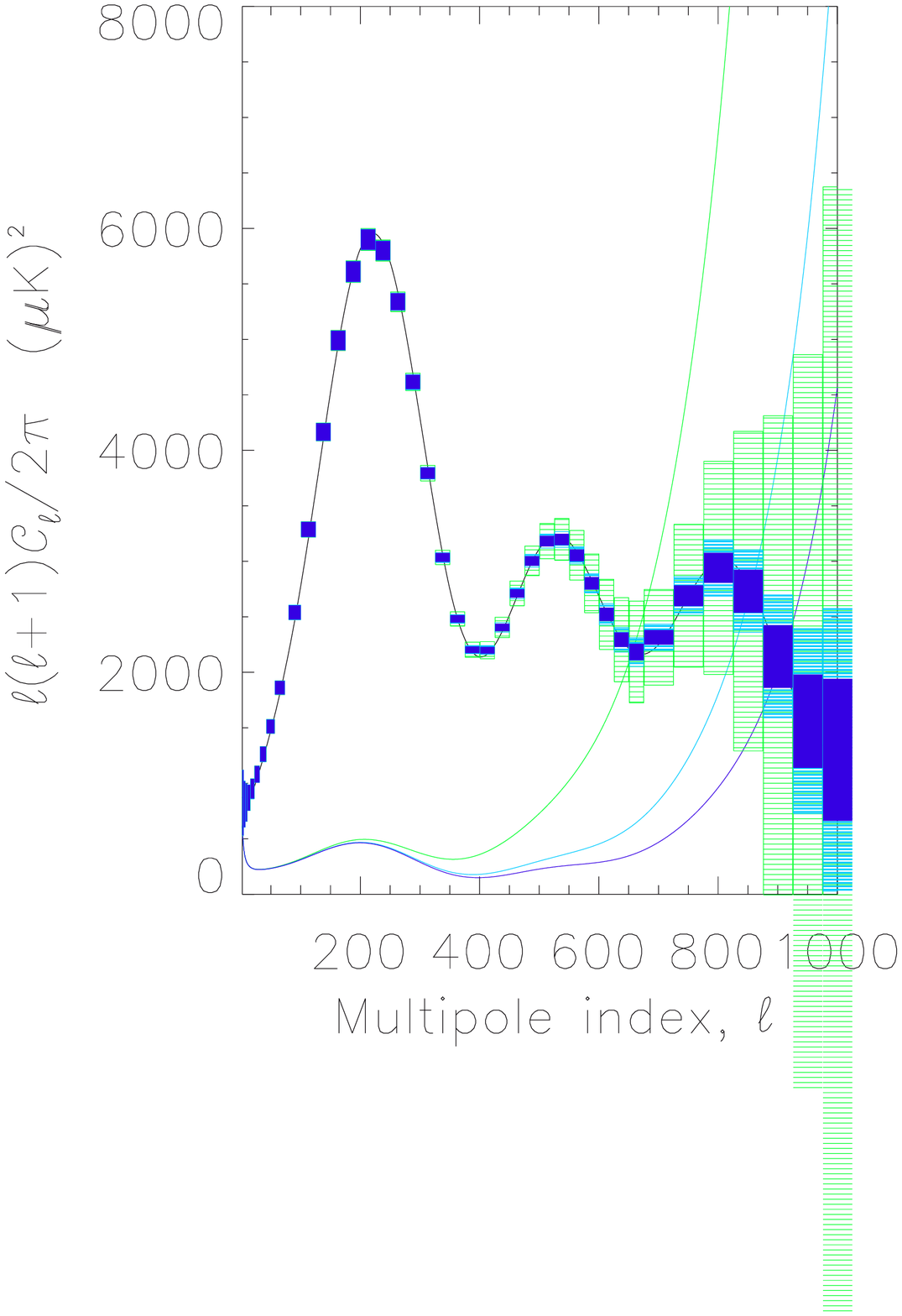}
  \end{center}
\caption{An estimate of the $1\sigma$ errors for a concordance
angular spectrum\cite{Wang00} based on Table 2. The smooth curves 
rising at high $l$
indicate the error per $l$; 
the width of a box indicates the window in $l$ over which the curve 
is averaged. The largest error bars are for the 6 month maps, the
next are for the 2 yr maps, and the narrowest are for the 4 yr maps. The
assumed sky coverage is 80\%.}
\end{figure}

\section{Acknowledgments}
Thank you for a wonderful, stimulating, and fun conference.
It was great. This work was supported by the NSF and NASA.
{\sl MAP} is a collaboration between 
NASA (Chuck Bennett [PI], Gary Hinshaw, Al Kogut, \& Ed Wollack), 
Princeton (Chris Barnes, Norm Jarosik, Michele Limon, Lyman Page, 
David Spergel \& David
Wilkinson), Chicago (Steve Meyer), UCLA (Ned Wright), UBC (Mark
Halpern), and Brown (Greg Tucker). {\sl MAP}'s success is the result of a
dedicated and incredibly hard working spacecraft team at NASA/GSFC 
led by the Project Manager Liz Citrin. More information may be found at
http://map.gsfc.nasa.gov
  

\end{document}